\documentclass[amsmath,preprint,letterpaper,prl,endfloats,11pt]{revtex4}
\usepackage{graphicx}
\usepackage{bm}
\usepackage{fancyhdr}
\usepackage{amsmath}
\usepackage{graphicx}
\usepackage{amsfonts}
\usepackage{amsbsy}
\usepackage{amssymb}

\begin{document}

\title{Heavy atom quantum diffraction by scattering from surfaces}
\author{Jeremy M. Moix}
\author{Eli Pollak}
\email{email: eli.pollak@weizmann.ac.il}
\affiliation{Chemical Physics Department, Weizmann Institute of Science, 76100 Rehovoth,
Israel}
\date{\today}

\begin{abstract}

Typically one expects that when a heavy particle collides with a surface,
the scattered angular distribution will follow classical mechanics. The
heavy mass assures that the de Broglie wavelength of the incident particle
in the direction of the propagation of the particle (the parallel direction)
will be much shorter than the characteristic lattice length of the surface,
thus leading to a classical description. Recent work on molecular
interferometry has shown that by increasing the perpendicular coherence 
length, one may observe interference of very heavy species passing through a
grating.  Here we show, using quantum mechanical simulations, that the same
effect will lead to quantum diffraction of heavy particles colliding with a
surface. We find that the effect is robust with respect to the incident
energy, the angle of incidence and the mass of the particle. It may also be
used to verify the quantum nature of the surface and its fluctuations at
very low temperatures.
\end{abstract}

\maketitle

\section{Introduction}

One of the fundamental building blocks of quantum mechanics is the wave
particle duality of matter. This duality was demonstrated convincingly
eighty years ago by Estermann and Stern when they reported the first
observation of atomic diffraction of He atoms scattered from a LiF surface~%
\cite{estermann30}. It took another forty years before the diffractive
scattering of Ne from LiF was measured by Williams~\cite{williams71}.
Subsequently Rieder and Stocker provided additional evidence for Ne
diffraction in scattering experiments with low index metal surfaces~\cite%
{rieder84}. Further progress was made when Schweizer and Rettner~\cite%
{schweizer89} reported the first observation of diffractive Ar scattering
from a hydrogen covered tungsten surface. However, the diffraction peaks
were observed only for rather low scattering energies and they were
superimposed on a broad background of ``classically" scattered Ar atoms.
Much sharper diffraction peaks were found over a decade later by Andersson
et al~\cite{andersson02} for the scattering of Ar and Kr from a Cu(111)
surface provided that the surface temperature was very low ($10$ K), that
the energy of the incident atom was low ($36$ meV) and that the angle of
incidence was large ($70^\circ$). However, in the majority of heavy atom
scattering experiments, the angular distribution is well described as
classical mechanical rainbow scattering~\cite{kleyn91}.

In parallel, especially during the past fifteen years there has been
significant progress made in the observation of diffraction for heavy atoms
and molecules when scattered through nanoscale gratings. Optics and
interferometry of atoms and molecules may be considered today as a highly
active area of research~\cite{cronin09}. As a striking example, we note the
interference pattern measured by Arndt et al~\cite{arndt99,nairz03} for the
fullerene molecule (C$_{60}$) scattered through a grating. By careful
velocity selection and the use of collimation slits, they could reduce the
uncertainty in the momentum perpendicular to the direction of propagation of
a molecular beam of fullerenes such that the wavelength of the center of
mass of the molecule in the perpendicular direction became larger than the
size of the grating ($\sim 100$ nm). They could then observe the
characteristic double slit diffraction pattern of the fullerene molecules as
they passed through the grating. These experiments have now led to the
observation of interference of other large organic molecules~\cite{gring10}.

The central theme of this letter is to replace the 100 nm scale grating with
the natural ``grating" of surfaces, where  the typical lattice length of a
metal, semiconductor or inorganic surface is of the order of 0.5-1 nm. One
should thus be able to use the velocity selection and collimation methods
employed in the previous experiments to create atomic and molecular beams
whose wavelength perpendicular to the direction of propagation of the
incident beam is of the order of 1 nm or longer in order to observe
diffraction of heavy atoms and molecules scattered from surfaces. The
resulting diffraction patterns may be obtained at essentially any incident
scattering angle and over a wide range of incident energies. The fact that
the \textquotedblleft grating size" is only 1 nm implies that the loss of
signal due to the distance from the source (leading to a quadratic loss of
signal strength) is orders of magnitude smaller than in experiments with a
100 nm grating.

The measurement of the diffraction pattern may also shed light on another
fundamental quantum property of matter, namely the low temperature zero
point energy fluctuations of surfaces. The diffraction pattern for heavy
particles is sensitive to the surface temperature. If the temperature 
is too high, the
fluctuations of the surface will smear out the diffraction pattern and one
will regain the classical rainbow scattering structure. The diffraction
peaks emerge as the temperature is lowered sufficiently such that the
thermal fluctuations of the surface atoms do not destroy the coherence of
the incident beam. This property implies that the diffraction pattern is
also a sensitive probe of the interaction of the heavy projectile with the
surface phonons and may be used to measure the interaction strength
(friction coefficient).

Finally, inspection of the two dimensional final momentum distribution of
the scattered beam reveals that the short wavelength of the incident beam in
the parallel direction (direction of propagation of the incident beam) is
observed as a smearing of the diffraction pattern in one direction. Instead
of the two dimensional diffraction pattern that is expected if the
wavelength of the incident beam is longer than the lattice length in both
the parallel and perpendicular directions, one observes in this case only a
one dimensional diffraction pattern reflecting the long wavelength in the
perpendicular direction only.

To demonstrate the diffraction of heavy atoms induced by long perpendicular
wavelengths, we will consider the in-plane scattering of Ar on a LiF(100)
surface which was measured extensively by Kondo et al~\cite{kondo06}. Their
experiments were carried out with the surface at room temperature and showed
the characteristic classical rainbow scattering pattern. The double peaked
distribution was observed over a relatively large range of incident energies
(from $315$~meV to $705$~meV) at an incidence angle of $45^{\circ }$. They
did not observe any hint of diffraction. Since the typical measurements are
for in-plane scattering, it suffices to model the system in terms of two
degrees of freedom. The vertical and horizontal coordinates of the incident
atom (with mass $M$) are denoted as $z$ and $x$, with conjugate momenta $%
p_{z}$ and $p_{x}$. 
The model Hamiltonian is then taken to have the form 
%\begin{equation}
%H_{\rm s} = \frac{p_{x}^{2}+p_{z}^{2}}{2M}
%+\bar{V}(z)+\bar{V}^{\prime }\left( z\right)
%h\sin \left( \frac{2\pi x}{l}\right) .  \label{1}
%\end{equation}%
\begin{equation}
H=\frac{p_{x}^{2}+p_{z}^{2}}{2M}+\bar{V}(z)+\bar{V}^{\prime }\left( z\right)
h\sin \left( \frac{2\pi x}{l}\right) .  \label{1}
\end{equation}%
The vertical potential $\bar{V}(z)$ is taken to be a Morse potential $\bar{V}%
(z)=V_{0}\left( 1-e^{-\alpha z}\right) ^{2}-V_{0}$ with $V_{0}$ the
physisorbed well depth and $\alpha $ the stiffness parameter. We assume a
sinusoidal surface corrugation with period $l$ (the lattice length) and
amplitude $h$. The corrugation is weak ($h/l\ll 1$), allowing for expansion
of the vertical potential in terms of a fluctuation along the horizontal
coordinate. This then leads to the coupling $\bar{V}^{\prime }\left(
z\right) $ between the vertical and horizontal modes of the incident atom.
The parameters used previously to model the experiment of Kondo et al were $%
\alpha l=5$, $\ $and $\alpha h=0.05$~\cite{moix10}. The physisorbed well
depth $V_{0}$ was taken to be $150$~meV (somewhat larger than the $88$~meV
of Ref.~\cite{klein79}) and the lattice length is $l=4$~\AA ~\cite{ekinci04}.

The incident argon wavepacket is initially sufficiently far removed from the
surface such that its interactions with the surface  vanish.  It is
described in terms of a Gaussian wavepacket. We use the notation $%
y_{\parallel }$ and $y_{\perp }$ to denote the coordinates of the wavepacket
in the parallel and perpendicular directions respectively. The associated
(positive) momenta are $p_{\parallel }$ and $p_{\perp }$ respectively. 
For
an incident scattering angle $\theta $ (taken to be negative) one has that $%
p_{z}=-p_{\parallel }\cos \theta $ and $p_{x}=-p_{\parallel }\sin \theta $.
The incident beam is thus described by the wave packet
\begin{eqnarray}
\Psi _{s}\left( y_{\parallel },y_{\perp };
  p_{\parallel },p_{\perp},y_{\parallel ,0},y_{\perp ,0}\right)  &=&
     \left( \frac{\Gamma _\parallel\Gamma _\perp}{\pi^2}\right)^{1/4} \notag \\
 && \exp \left( -\frac{
  \Gamma _{\perp }\left( y_{\perp }-y_{\perp ,0}\right) ^{2}
 +\Gamma _{\parallel }\left( y_{\parallel }-y_{\parallel ,0}\right)^{2}
 } {2} \right)   \notag \\
 &&\exp \left( \frac{i}{\hbar }\left[ 
  p_{\perp }\left(y_{\perp}-y_{\perp,0} \right) 
 -p_{\parallel }\left( y_{\parallel}-y_{\parallel ,0}\right) 
\right] \right) .  \label{2}
\end{eqnarray}

Of central importance are the respective width parameters $\Gamma
_{\parallel}$ and $\Gamma _{\perp}$. In the experiments of Kondo et al, the
uncertainty in the energy of the incident argon beam is reported as 
$\Delta E/E_i = 17.7\%$ for $E_i = 315$ meV and increasing to 
$\Delta E/E_i = 26.9\%$ at $E_i = 705$ meV~\cite{kondo06}. 
From the energy-momentum relationship we have that 
$\Delta p/p_i = \frac{\Delta E}{2E_i}$, this implies that 
$l\sqrt{ \Gamma _{\parallel }}=39\mbox{ }(88)$ at $E_i = 315\mbox{ }(715)$ 
meV. The experimental paper does not provide us with the angular width of
the incident beam.
For $E_i=705$ meV, an angular width of $2^\circ$ implies that 
$l\sqrt{\Gamma_{\perp}}=17$. 
These choices for the width parameters result in a wavepacket that is localized
in configuration space as compared with the lattice length.

The final momenta distribution can be obtained from the formally exact
expression~\cite{moix09},
\begin{equation}
P\left( p_{x_{f}},p_{z_{f}};t\right) =
\frac{1}{l}\int_{0}^{l}dx_{0} %\int_{-\infty }^{\infty }dp_{b_{f}}\times 
\left\vert \left\langle
p_{x_{f}},p_{z_{f}}\left\vert \exp \left( -\frac{i}{\hbar }\hat{H}t\right)
\right\vert \Psi _{s}\left( p_{x_{0}},p_{z_{0}},x_{0},z_{0}\right)
\right\rangle \right\vert ^{2}.  \label{3}
\end{equation}%
The corresponding angular distribution is obtained from a simple
transformation of the momenta distribution to radial coordinates followed by
integration over the radial momentum~\cite{moix09}. The propagation time $t$
is taken to be sufficiently long so that the scattered wavepacket is out of
the range of interaction with the surface.

\begin{figure}
\includegraphics*[width=0.85\textwidth]{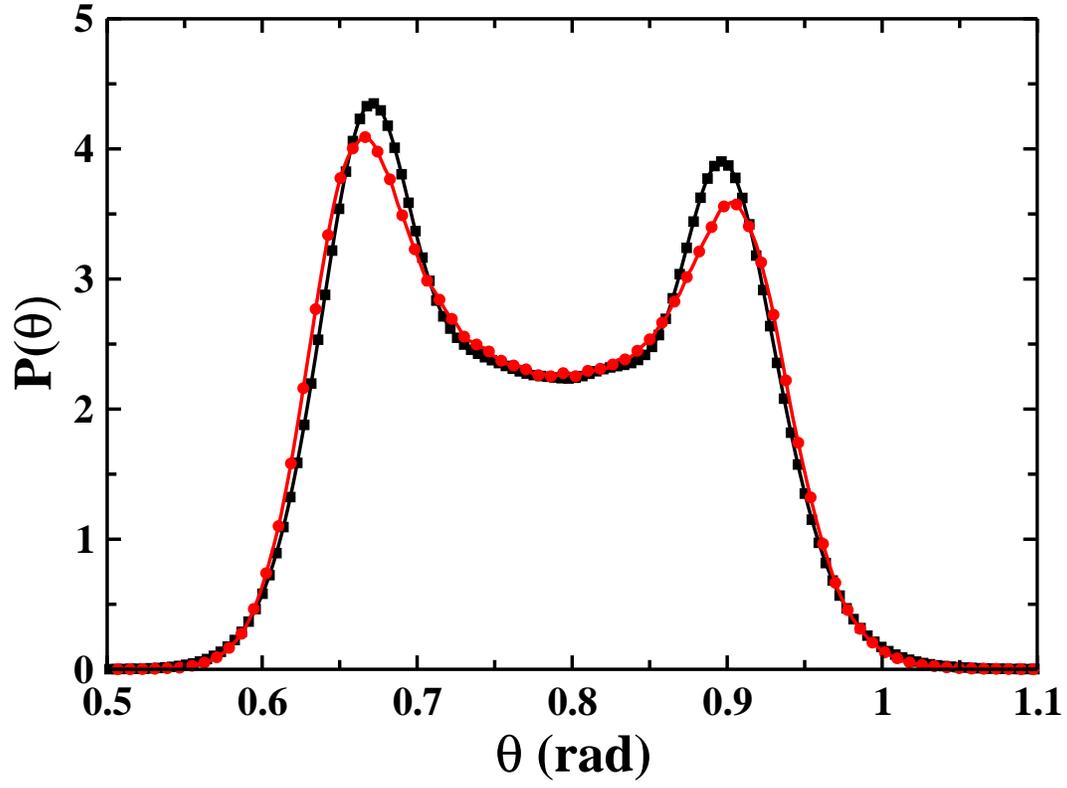}
\caption{ 
Final angular distributions at an incident energy of $705$ meV and 
incident angle of $\pi/4$.
The classical Wigner result is depicted as the solid (red) line with
dots and the quantum mechanical result is shown as the solid (black) line
with squares. }
\label{fig:fig1}
\end{figure}

\begin{figure}
\includegraphics*[width=0.85\textwidth]{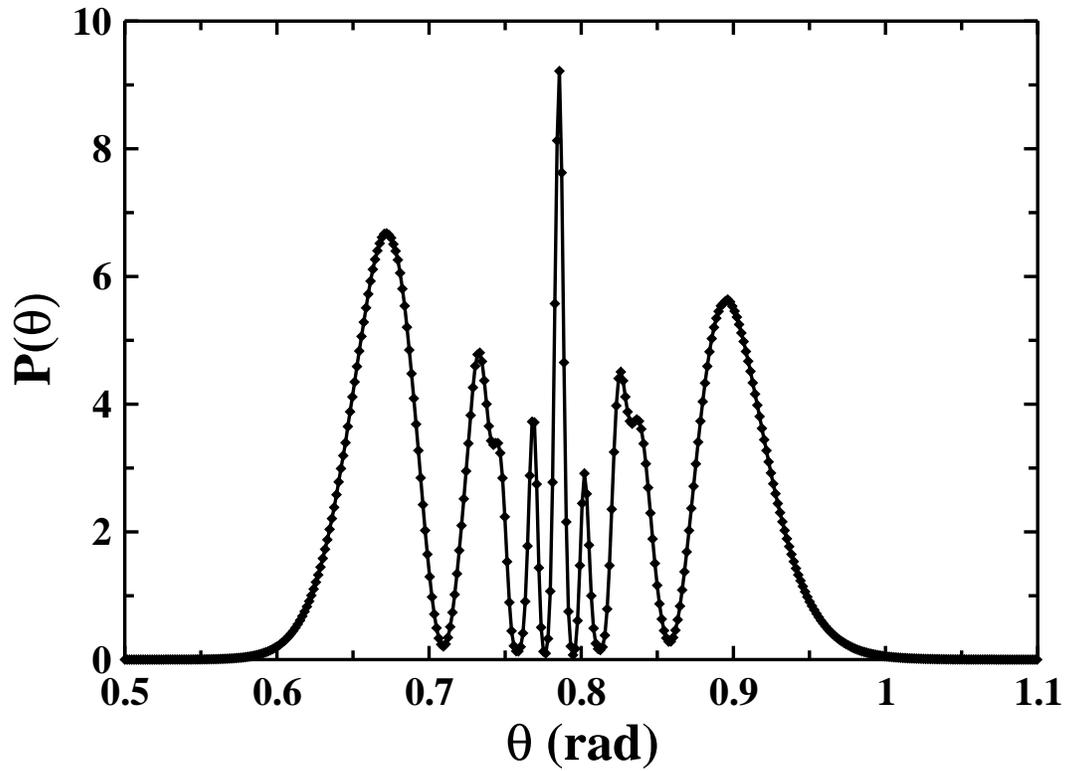}
\caption{ 
Final angular distribution for argon scattering at an
incident energy of $E_i = 705$ meV and incident angle of $\pi/4$ but with 
a perpendicular de Broglie wavelength which is much larger than the lattice
length.  
Note the resulting diffraction pattern.}
\label{fig:fig2}
\end{figure}

In Fig.~\ref{fig:fig1} we plot the angular distributions at an incidence energy of $E_i=705$ meV. 
The width parameters are taken as described above, 
using the experimental uncertainty in the incident
beam characteristics and an angular width of $2^{\circ}$.  
The angle of incidence is $45^{\circ }$. 
The wavepacket propagation was carried out using the split operator method with
a grid size of $1024 \times 2048$ for the horizontal and vertical
coordinates respectively. The results displayed no change with respect to
doubling the grid sizes. 
The average over the unit cell in Eq.~\ref{3} was evaluated from 20 different
impact parameters. 
The quantum distribution is compared to the classical Wigner angular 
distributions as defined in Eq.~2.7 of Ref.~\cite{pollak09}. 
It is evident that due to the
relatively large uncertainty in the velocity of the incident beam, one
observes only the classical scattering distribution; quantum effects are
washed out. The central result of this paper is shown in Fig.~\ref{fig:fig2}. 
Reducing the perpendicular width parameter to the value of 
$l\sqrt{\Gamma_{\perp }} =1.6$ while leaving the parallel width parameter
unchanged gives the quantum momentum distribution shown in the figure. 
The diffraction pattern is now evident, even though the incident atom is argon. 
We have carried out computations at a lower energy of 315 meV 
which display similar diffractive features to those seen in Fig.~\ref{fig:fig2}.
Changing the angle of incidence also does not destroy the quantum
diffraction pattern. 
As long as the coherence length in the perpendicular direction is longer than
the lattice length, one observes diffraction.  In this paper we concentrated
on the case of Ar scattering due to the experimental data available for
this case which provides us with a reasonable model and a prediction which
should be readily borne out by further experiment.
However, there is no reason to believe that the results would change
if we substitute Ar for a heavier atom such as Rubidium.

\begin{figure}
\includegraphics*[width=0.85\textwidth]{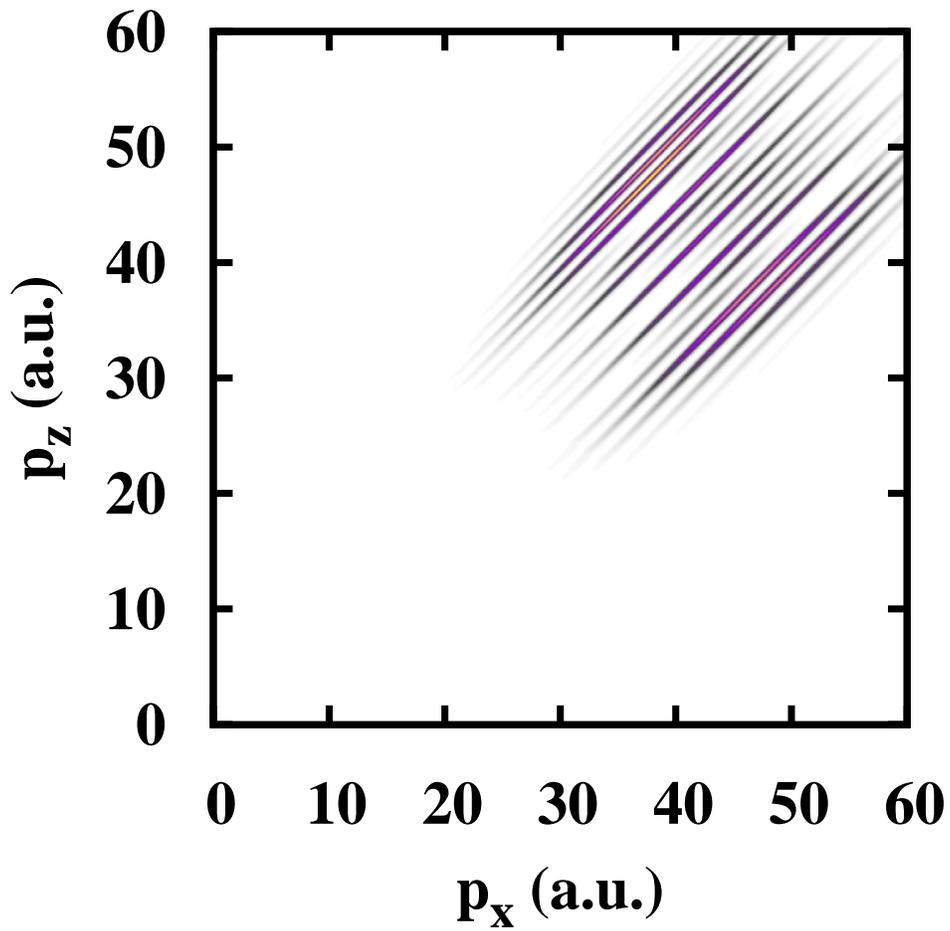}
\caption{Final momentum distribution for argon scattering at an
incident energy of $E_i = 705$ meV and incident angle of $\pi/4$ with the width
parameter of $l \sqrt{\Gamma_{\perp }} = 1.6$. 
Note the ``cigar" shape of each separate peak.}
\label{fig:fig3}
\end{figure}

An interesting aspect of this type of preparation is the change in the
momentum distribution which is shown in Fig.~\ref{fig:fig3} 
for an incident energy $E_i=705$ meV.
When the de Broglie wavelength is sufficiently longer than the lattice length
in both the parallel and perpendicular directions, 
one will observe that the momentum distribution is comprised
of isolated delta functions spread out in the momentum plane. 
However, here, one de Broglie wavelength is long whereas the other is short. 
As a result, the momentum
distribution takes the form of a collection of cigar shaped distributions,
where one direction is ``discretized" and the other remains continuous. 
It is the discretized direction which gives rise to the angular diffraction
pattern.

In summary, using model computations we have shown that collimation of a
beam of atoms in the direction perpendicular to the direction of propagation
may lead to a distinctive diffraction pattern.
Diffraction is observed irrespective of the incident energy and angle 
even though the mass of the atom is large and the
natural de Broglie wavelength associated with the beam is of the order of
picometers. 
This demonstrates that one should be able to observe heavy
particle diffraction in the scattering of atoms and molecules from surfaces
under experimental conditions that are much less severe than those needed
for coherent scattering through 100 nm gratings. On the down side,
decoherence through collision with a surface will be greater than when
traversing through a grating. Both the interaction with surface modes as
well as energy exchange between internal degrees of freedom may lead to
decoherence. On the other hand, the observation of such decoherence may lead
to important information about the internal modes of the molecule as well as
the surface fluctuations which induce such decoherence. 
These topics are part of an ongoing study.

\textbf{Acknowledgment} This work was supported by grants from the Israel
Science Foundation and the Weizmann-UK Joint Research Program.

\newpage

\end{document}